\documentclass[runningheads]{llncs}
\usepackage{graphicx}
\usepackage{tikz}
\usepackage{booktabs}
\usepackage{comment}
\usepackage{amsmath,amssymb} % define this before the line numbering.
\usepackage{color}
\usepackage{orcidlink}
\usepackage{cleveref}

\usepackage[accsupp]{axessibility}  % Improves PDF readability for those with disabilities.

\begin{document}
% \linenumbers
\pagestyle{headings}
\mainmatter
\def\ECCVSubNumber{6}  % Insert your submission number here

\title{ReLaX: Retinal Layer Attribution for Guided Explanations of Automated Optical Coherence Tomography Classification} % Replace with your title

\begin{comment}
\titlerunning{ECCV-22 submission ID \ECCVSubNumber} 
\authorrunning{ECCV-22 submission ID \ECCVSubNumber} 
\author{Anonymous ECCV submission}
\institute{Paper ID \ECCVSubNumber}
\end{comment}
%******************

%\begin{comment}
\titlerunning{Retinal Layer Attribution for Guided Explanations of OCT Classification}
\author{Evan Wen\inst{1}\orcidlink{0000-0003-4461-4584} \and
ReBecca Sorenson\inst{2} \and
Max Ehrlich\inst{3,4}}
\authorrunning{E. Wen et al.}
\institute{The Pingry School, Basking Ridge, NJ\\
\email{ewen2023@pingry.org}\and
Department of Optometry, Chinle Comprehensive Health Care Facility, Chinle, AZ\and
Department of Computer Science, University of Maryland, College Park, MD\and
NVIDIA, Santa Clara, CA}
%\end{comment}
%******************
\maketitle
  
\begin{abstract}
    30  million  Optical  Coherence  Tomography  (OCT)  imaging tests are issued annually to diagnose various retinal diseases, but accurate diagnosis of OCT scans requires trained eye care professionals who are still prone to making errors. With better systems for diagnosis,  many cases of vision loss caused by retinal disease could be entirely avoided.  In this work, we present ReLaX, a novel deep learning framework for explainable, accurate classification of retinal pathologies which achieves state-of-the-art accuracy. Furthermore, we emphasize producing both qualitative and quantitative explanations of the model’s decisions. While previous works use pixel-level attribution methods for generating model explanations, our work uses a novel retinal layer attribution method for producing rich qualitative and quantitative model explanations. ReLaX determines the importance of each retinal layer by combining heatmaps with an OCT segmentation model. Our work is the first to produce detailed quantitative explanations of a model’s predictions in this way. The combination of accuracy and interpretability can be clinically applied for accessible, high-quality patient care.

\dots
\end{abstract}

\section{Introduction}

Every year there are approximately 30 million Optical Coherence Tomography (OCT) procedures done worldwide \cite{pmid27409459}. OCT is a non-invasive imaging test that yields cross-sectional slices of a patient's retina \cite{pmid10933065} which can be used to diagnose a multitude of retinal diseases. It is estimated that up to 11 million people in the United States have some form of macular degeneration \cite{brightfocusfoundation_2020}. The number of retinal disease patients continues to increase making the need for accurate diagnosis ever so more important. Accurate OCT diagnosis has traditionally been done by eye care professionals trained to interpret these scans. However, the surplus of patients has been met with a shortage of eye care professionals leading to occurrences of potentially unnecessary cases of vision loss \cite{pmid28128796}.

\begin{figure*}[t]
    \centering
    \includegraphics[width=\textwidth]{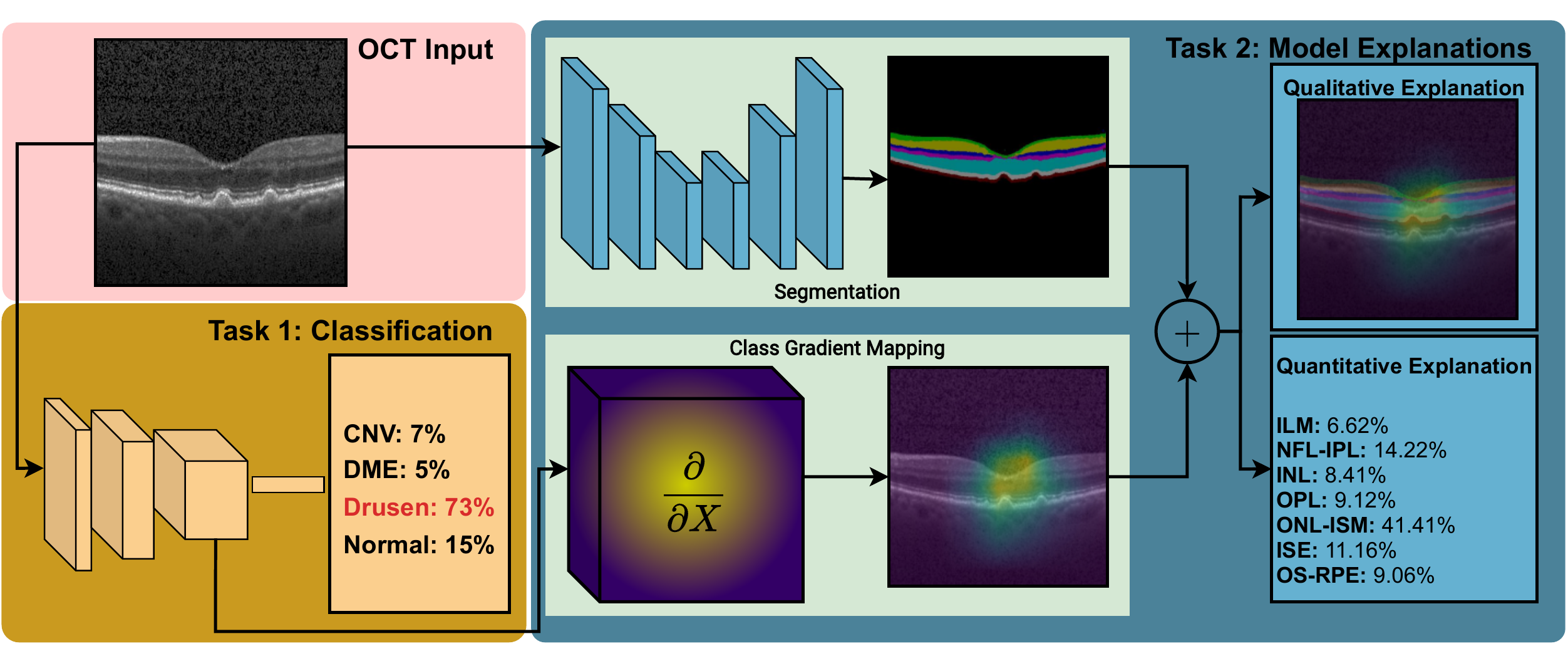}
    \caption{A diagram of ReLaX, a framework for generating accurate, interpretable OCT scan classifications. }
    \label{fig:overview}
\end{figure*}

In order to facilitate a faster diagnosis, many works have used deep learning to analyze retinal imaging. Used correctly, machine learning has the potential to reduce the load on ophthalmic clinics. A common flaw of these models is the lack of interpretability: they are black-boxes. To improve retinal disease patient care, machine learning models must be accurate but also more interpretable. A standalone diagnosis can leave patients skeptical of the model's validity. Another potential benefit of interpretability is when a model's interpretation is statistically suspicious, it can be flagged for review by a doctor for an extra level of validation, this requires \textbf{quantitative} as well as \textbf{qualitative} explanations. Previous works \cite{kermany2018identifying} have attempted to bring a degree of interpretability through attribution based methods such as occlusion testing. However, these approaches are insufficient for a clinical setting. These methods can provide a general idea of the model's interpretation of an image but do not obtain the same meticulousness as a human eye care professional.

ReLaX aims to provide more clear explanations through the use of retinal layer attribution. We present an accurate and interpretable deep learning framework for classifying OCT scans. Retinal layer attribution uses a unique combination of Gradient Weighted Class Activation Mapping (GradCAM) \cite{DBLP:journals/corr/SelvarajuDVCPB16} and semantic segmentation to obtain detailed breakdowns of exactly which retinal layers the classification model uses when making its decisions. To our knowledge, we are the first to incorporate the use of segmentation maps for explaining the predictions of a classifier. This allows us to understand the specific behaviors the model learns and analyze why the model makes incorrect decisions. Our method produces qualitative explanations, in the form of heatmaps with highlighted retinal layers and quantitative explanations indicating the percent of model focus on each retinal layer. ReLaX can also go beyond providing explanations by aiding clinical diagnoses through error analysis. Furthermore, our method is completely model agnostic meaning that any CNN classifier and segmentation model are acceptable. The algorithm isn't limited to strictly CNN architectures either as the GradCAM heatmaps can be substituted with attention maps from a ViT architecture \cite{DBLP:journals/corr/abs-2010-11929}. With further analysis, we are able to show that deep learning models interpret medical images in a similar way to human professionals. This proposed approach is evaluated on a publicly available OCT data set. See Figure \ref{fig:overview} for an overview of our algorithm. In summary, we contribute:
\begin{itemize}
    \item A CNN architecture that achieves state-of-the-art results in OCT classification.
    \item A novel \textbf{retinal layer attribution} concept 
    \item A novel method for producing both \textbf{quantitative} and \textbf{qualitative} explanations of the machine generated diagnosis.
\end{itemize}

\section{Literature Review}
\subsection{OCT Classification}
\label{2.1}
Due to the shortage of experienced eye care professionals and subjectivity in OCT classification, researchers have attempted to apply machine learning algorithms, most notably various types of Convolutional Neural Networks (CNN)~\cite{NIPS2012_4824}. Fauw et al.~\cite{de_fauw} used a 3D U-Net for segmentation combined with a 3D CNN to classify Normal, Choroidal Neovascularization (CNV), Macular Retinal Edema, Full Macular Hole, Partial Macular Hole, Central Serous Retinopathy(CSR), and Geographic Atrophy . Their models reached or exceeded the performance of human experts. Lu et al.~\cite{10.1167/tvst.7.6.41} used ResNet-101 to classify Cystoid Macular Edema, Epiretinal Membrane, Macular Hole, and Serous Macular Detachment, their model outperformed two physicians. Yoo et al.~\cite{pmid33492598} used Inception-V3 combined with CycleGAN for data augmentation to classify rarer retinal diseases. Nagasato et al.~\cite{pmid31697697} used a deep convolutional neural network to detect a nonperfusion area in OCTA images. Wang et al.~\cite{8794616} used CliqueNet to classify Age-related Macular Edema (AMD), DME, and Normal. Li et al.~\cite{pmid31853395} used ResNet50 to classify CNV, Diabetic Macular Edema (DME), Drusen, and Normal. Tsuji et al.~\cite{pmid32192460} used a capsule network to classify the same set of diseases. Also on the same data set were the works of Kumar et al.~\cite{puneet_optical_2022} and Asif et al.~\cite{asif_amjad_qurrat-ul-ain_2022}. Kumar et al.~\cite{puneet_optical_2022} utilized the concept of attention in addition to a deep CNN for OCT classification. Aisf et al.~\cite{asif_amjad_qurrat-ul-ain_2022} used an architecture similar to ResNet50 with a transfer learning approach to classify the scans. Saleh et al.~\cite{saleh_transfer_2022} used an Inception V3 Net for the same task. Of the works listed so far, many have already achieved high enough accuracy for clinical implementation. Most notably Tsuji et al.~\cite{pmid32192460} achieved a classification accuracy of 0.996. The primary drawback of these deep learning solutions is that they operate within a "black box" making it difficult to understand the model's decision-making process. Our work  contributes a new CNN architecture which achieves state-of-the-art performance for OCT classification. Furthermore, we achieve a new level of detail in producing model explanations. Previous works were limited to qualitative explanations in heatmaps. However, heatmaps have little to no use when one cannot understand the specific parts of the retina the model is looking at. Due to this constraint, we develop a segmentation-based algorithm for generating quantitative explanations. Our algorithm not only provides novel quantitative explanations but also richer, more in-depth qualitative explanations.  

\subsection{CNN Visualization}
Various approaches have been used to visualize the behaviours of CNNs. In Zeiler et al, a deconvolutional network was used to map network activations to the input pixel space and show the input patterns learned by the CNN \cite{DBLP:journals/corr/ZeilerF13}. Simonyan et al. visualized partial derivatives from predicted class scores and generated saliency maps \cite{simonyan_vedaldi_zisserman_2014}. Springenberg et al. used a deconvolution approach similar to Zeiler et al. for feature visualization of a CNN \cite{1412.6806}. Zhou et al. introduced class activation mapping (CAM), a technique for generating localization maps using the global average pooled convolutional feature maps in the last convolutional layer \cite{zhou2015cnnlocalization}. Selvaraju et al. generalized the CAM algorithm by making GradCAM, a technique that does not require a specific CNN architecture to develop explanations. CAM, however, requires a CNN with global average pooling followed by a single fully connected layer that outputs the prediction \cite{DBLP:journals/corr/SelvarajuDVCPB16}. GradCAM first computes the gradient of the targeted output class with respect to the final convolutional layer's feature map activations. Next, it performs pooling on the gradients to obtain neuron importance weights before performing a weighted combination of the feature maps to compute the GradCAM heatmap. These methods are considered forms of \textbf{attribution} as they highlight areas of the input that contribute most to the classification.  

\subsection{Attempts at Explainable OCT Classification} 

Of the works listed in Section \ref{2.1}, some have used attribution-based methods to generate heatmaps from model predictions. Tsuji et al.~\cite{pmid32192460} generated heatmaps using an algorithm inspired by CAM~\cite{zhou2015cnnlocalization}. In their work, expert ophthalmologists assessed the heatmap images and confirmed that the activated regions were the correct regions of focus and thus the model was trained accurately. They also performed some error analysis by analyzing heatmaps from their incorrect predictions. However, they note that in wrong predictions the heatmaps still show that the model looked in the correct area. Similarly, Kermany et al.~\cite{kermany2018identifying} used occlusion testing to generate heatmaps which they also confirmed with human experts. One issue with this form of explanations is that the explanations have little variance between correct and incorrect classifications. How can one estimate the certainty of prediction in a clinical setting when the explanations are always similar? Furthermore, the qualitative explanations are not very robust. The human professionals do seem to agree with the models' areas of focus, however, given that the heatmaps are relatively coarse and human classification can be subjective, the qualitative explanations can still be dubious and difficult to trust. Therefore, the objectives of our proposed algorithm are to 
\begin{itemize}
    \item Demonstrate a stronger connection between human professional examinations and machine learning model interpretations
    \item More precisely pinpoint the areas being focused on by the model
    \item Perform meaningful error analysis 
\end{itemize}
\section{Methodology}

Here we describe ReLaX, our algorithm to produce highly accurate, easily understandable diagnoses. Our algorithm depends on a classification model, GradCAM heatmaps, and a segmentation model which are now discussed in more detail. The motivation of our approach is that OCT scans show each of the retinal layers. Instead of thinking about the attribution-based method in the context of the whole image, we can focus on highlighting the importance of the retinal layers rather than of the individual pixels. 

\subsection{Classification Model}

For OCT image classification we build a CNN architecture based off an EfficientNetB2 backbone~\cite{DBLP:journals/corr/abs-1905-11946}. We choose the EfficientNet model because of its robust performance on the ImageNet-1k data set \cite{deng2009imagenet}. Since EfficientNet is designed for ImageNet classification, it outputs 1000 logits in the final layer. Therefore, we add two fully connected layers of sizes 100 and 4 with a softmax activation on the final layer to perform classification on four retinal diseases.

The full model is trained end-to-end with the Adam \cite{Kingma2015AdamAM} Optimizer with a learning rate of 0.001 with the categorical cross-entropy loss function. We train the model for a total of 20 epochs. We use four methods for evaluating our models: accuracy, precision, recall, and F1 score.

\subsection{Gradient-Weighted Class Activation Mapping}
We use Gradient-weighted Class Activation Mapping (GradCAM) \cite{DBLP:journals/corr/SelvarajuDVCPB16} to produce heatmaps highlighting the regions the model utilized to make a prediction. The GradCAM algorithm utilizes the inputs to the model, the output of the final convolutional layer, and the output of the model prior to the softmax activation function. We overlay heatmaps on the original images to compare the model's focus during classification of each disease. We also analyze why the model makes incorrect classifications by looking at its resulting heatmap. Standalone GradCAM heatmaps can give a level of interpretability similar to previous works: a visual sense of the regions of focus in the image. 

\subsection{Segmentation Model}

To fully implement the idea of retinal layer attribution, we employ a U-Net \cite{DBLP:journals/corr/RonnebergerFB15} architecture, pretrained for retinal layer segmentation \cite{DBLP:journals/corr/RoyCKSKWN17}. Standalone GradCAM heatmaps give a general idea of where the model looked at, but by localizing the heatmap to specific retinal layers, we obtain both a clearer qualitative explanation of the model as well as a thorough quantitative explanation. The U-Net is trained to detect nine retinal layers: region above retina (RaR), inner limiting membrane (ILM), nerve fiber ending to inner plexiform layer (NFL-IPL), inner nuclear layer (INL), outer plexiform layer (OPL), outer nuclear layer to inner segment myeloid (ONL-ISM), inner segment ellipsoid (ISM), outer segment to retinal pigment epithelium (OS-RPE), and region below retina (RbR). The U-Net model is trained on DME scans from the data set presented in Srinivasan et al. \cite{Srinivasan:14} for 20 epochs.

\subsection{Calculating Retinal Layer Attribution}

The full proposed ReLaX method consists of applying the segmentation model on our OCT data set before overlaying the GradCAM heatmaps on the segmentation maps. This allows us to obtain a visual breakdown of the model's focus based on the retinal layers. We also calculate percentages of the model's focus on each of the retinal regions excluding the regions above and below the retina. We choose to exclude these regions because they generally are less important for the model, but they have high enough area to potentially alter the percentages of the model's focus. For each of the four retinal disease classifications, we obtain an average retinal layer focus percentage for the seven retinal layers, denoted in \Cref{eq:Fi}, during correct OCT classifications. $R_i$ denotes the model's focus on layer $i$. $S_{i,r,c}$ is a one-hot encoded integer as to whether the pixel at ${r,c}$ is of retinal layer $i$, and $H_{r,c}$ is the value of the GradCAM heatmap at pixel ${r,c}$.

\begin{equation}
    \centering 
    \label{eq:Fi}
    R_i = 100 * \frac{{\sum_{r,c}^{}{S_{i,r,c}H_{r,c}}}}{\sum_{l=1}^{7}{\sum_{r,c}^{}{S_{l,r,c}H_{r,c}}}}
\end{equation}

\section{Results}

\subsection{Data Set}

The OCT scans for this work come from the Kermany labeled OCT data set \cite{kermany2018identifying}. The data set contains four classes: Choroidal Neovascularization (CNV), Diabetic Macular Edema (DME), Drusen, and Normal (healthy). The data set utilizes Spectralis OCT machine scans from the Shiley Eye Institute of the University of California San Diego, the California Retinal Research Foundation, the Medical Center Ophthalmology Associates, the Shanghai First People's Hospital, and the Beijing Tongren Eye Center. All images went through an extensive manual classification process by a cohort of retinal specialists to ensure correct labels.

We test class-weighting due to the imbalance in the data set. Drusen had significantly less scans in the training set whereas CNV had significantly more scans. The class weights are evaluated based on the number of scans in the training set. We take the highest number of scans for any label which was 37,205 for CNV and divide by the number of scans for each of the labels to obtain the class weights of 1, 3.279, 4.318, and 1.414 for CNV, DME, Drusen, and Normal respectively.

\subsection{Data Preprocessing}
The Kermany data set contains a total of 84,484 OCT scans. The data is split into 968 test images and 83,484 training images. Of the 83,484 training images, 37,205 are labeled as CNV, 11,348 are DME, 8616 are Drusen, and 26,315 are Normal. The test data consists of 242 images of each class. Scans are separated by their correct labels into folders containing JPEG image files. This choice of data split was specified by Kermany et al in the original work and has since been adopted for other recent works. All works in \Cref{Tab:metrics} are evaluated on the same data split. The dimensions of the original images vary from 384-1536 pixels wide and 496-512 pixels high. All images are rescaled to 260 pixels wide and 260 pixels high using bilinear interpolation. The training data is processed in batches of size 64.
\begin{table}[t]
    \centering
    \caption{Results of our model measured in accuracy, precision, recall, and F1 score. A comparison with previous works is shown; the bold represents the highest scores for each metric. Results shown as reported in previous works. }
    \begin{tabular}{lllll}
        \toprule
        Model                        & Accuracy        & Precision       & Recall          & F1-Score        \\\midrule
        \textbf{Ours}                & \textbf{0.998} & \textbf{0.998} & \textbf{0.998} & \textbf{0.998} \\
        Kermany et al. Inception V3  & 0.961           & 0.96125         & 0.961           & 0.963          \\
        Tsuji et al. Capsule Network & 0.996           & 0.996           & 0.996          & 0.998          \\
        Asif et al. Residual Network & 0.995           & 0.995           & 0.996          & 0.995           \\
        Kumar et al. Deep CNN       & 0.956           & -              & -           & -             \\
        Saleh et al. Inception V3 & 0.984           & -                & -           & -            \\
        Li et al. ResNet-50          & 0.973           & 0.963           & 0.985           & 0.975           \\
        \bottomrule
    \end{tabular}
    \label{Tab:metrics}
\end{table}
\begin{figure}[t]
    \centering
    \includegraphics[width=0.4\linewidth]{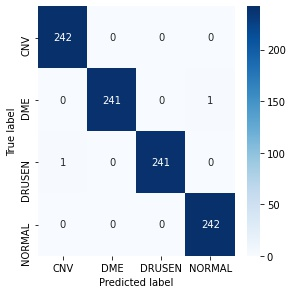}
    \caption{Confusion matrix from our classification model.}
    \label{fig:confmat}
\end{figure}
\subsection{Results of the Classification Models}

\Cref{Tab:metrics} shows the state-of-the-art performance of our classification model. The per-class performance of the model is shown in \Cref{fig:confmat}. As seen in \Cref{fig:confmat}, the model makes only two errors: one mistaking Drusen for CNV and one mistaking DME for Normal. The model performs more accurately than the models presented by Kermany et al. \cite{kermany2018identifying}, Tsuji et al. \cite{pmid32192460}, and Li et al \cite{pmid31853395} who were trained to classify the same four diseases.

\begin{figure}[t]
    \centering
    \includegraphics[width=0.67\linewidth]{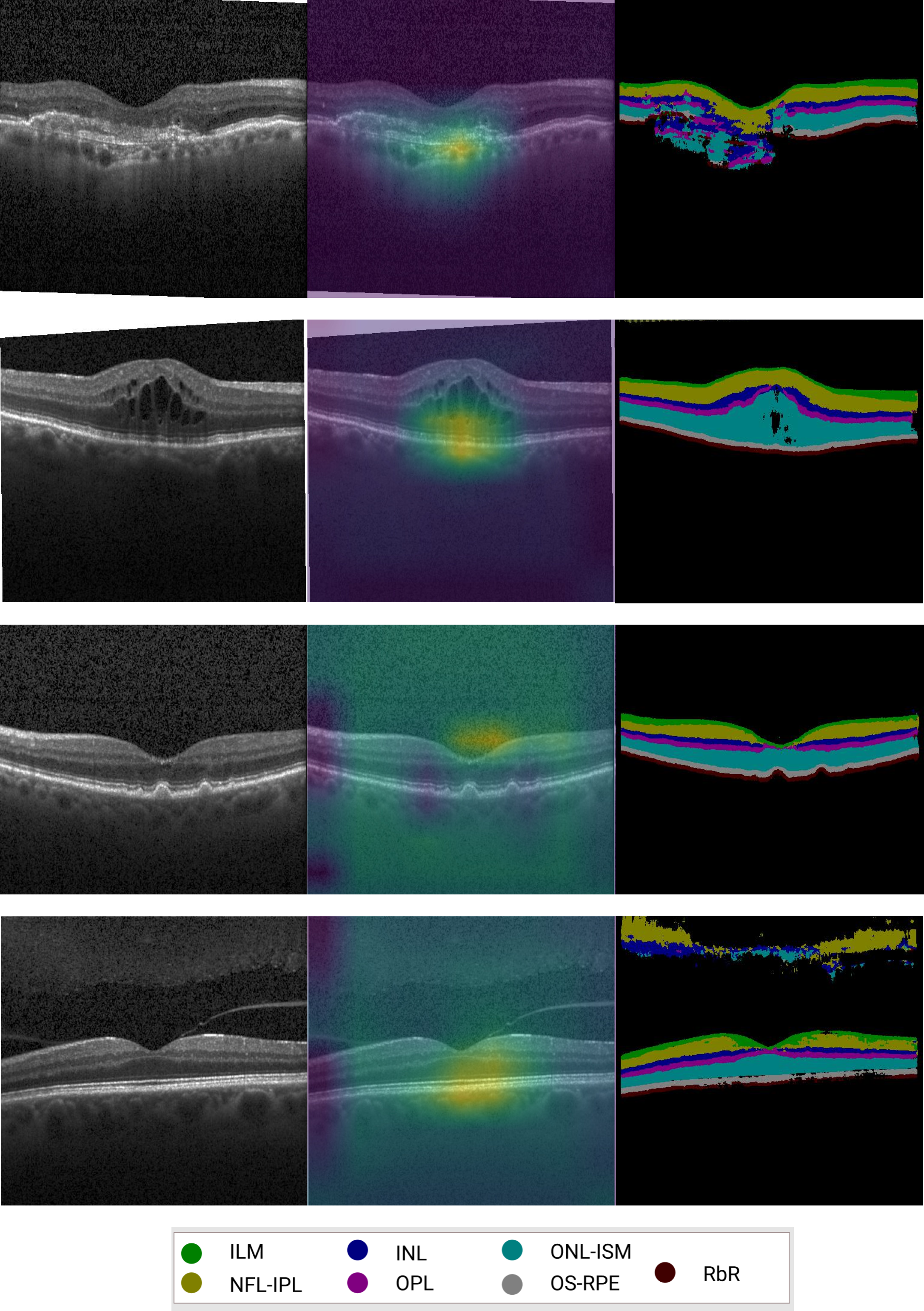}
    \caption{Heatmaps and segmentation maps from the OCT scans. Left: OCT scan, Middle: GradCam heatmap, Right: segmentation map. The scans are CNV, DME, Drusen, and Normal from top to bottom.}
    \label{fig:seg1}
\end{figure}
\subsection{Preliminary Qualitative Interpretability Results}
\Cref{fig:seg1} shows GradCAM heatmaps and segmented OCT scans of each classification. Heatmaps from the Drusen and Normal scans have less centralized focus, but they still emphasize the center of the scan. In the CNV classification, the model has the highest focus on the bottom-most layer of the retina. In the DME classification, the model still focuses on the bottom layer but also focuses on the intraretinal fluid. In the Drusen classification, the model does not have as much central focus. In the Normal classification, the model has more central focus than the Drusen classification. The model looks at the center of the scan to make the Normal classification.

The U-Net model performs well with segmentation on DME, Drusen, and Normal. The model is able to accurately detect each of the retinal layers. The model, however, does have some confusion on the region above the retina on the Normal scan. The scan appears to have some noise in the upper region which is captured in the segmented scan. The CNV scan is the least accurate of the segmented scans. The model has significant difficulty detecting the retinal layers in the irregular subretinal region. The model's difficulty with the CNV and Normal scans are likely due to the U-Net being unfamiliar with these type of scans. Given that the U-Net was trained only on DME scans, it makes sense that the U-Net does not perform as well on CNV and Normal scans. We explain how the inaccuracies in segmentation affect the interpretability algorithm in the next section. 

\begin{table}[t]
    \centering
    \caption{The model's mean focus on each of the retinal layers when making correct classifications. Each column provides a breakdown of the model's focus for correct classifications of each disease.}
    \begin{tabular}{ccccc}
        \toprule
        Layer   & CNV     & DME     & Drusen  & Normal  \\\midrule
        ILM     & 5.39\%  & 6.62\%  & 10.87\% & 7.96\%  \\
        NFL-IPL & 14.03\% & 14.22\% & 23.67\% & 20.48\% \\
        INL     & 12.26\% & 8.41\%  & 10.00\% & 8.75\%  \\
        OPL     & 12.73\% & 9.12\%  & 10.70\% & 9.97\%  \\
        ONL-ISM & 35.90\% & 41.41\% & 27.74\% & 30.91\% \\
        ISE     & 9.90\%  & 11.16\% & 9.61\%  & 13.18\% \\
        OS-RPE  & 9.78\%  & 9.06\%  & 7.42\%  & 8.75\%  \\
        \bottomrule
    \end{tabular}

    \label{tab:seg2}
\end{table}

\begin{figure}[t]
    \centering
    \includegraphics[width=0.73\linewidth]{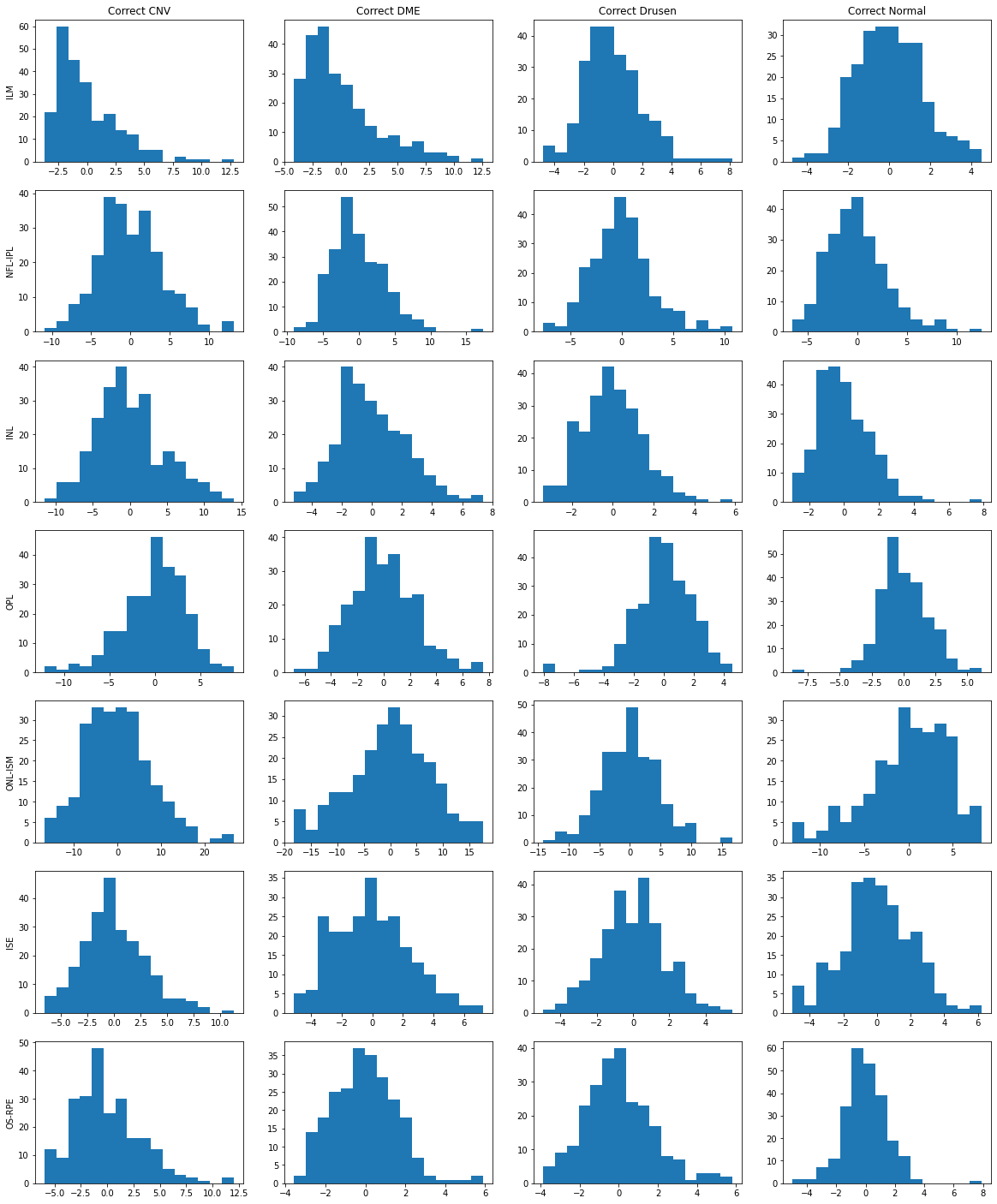}
    \caption{Histograms consisting the numeric deviation from the mean during correct classifications (e.g. a zero value means that the layer involvement for a particular correct classification has the exact same value as the mean layer involvement for that particular classification). Columns: The Pathology Being Classified. Rows: The Retinal Layers}
    \label{fig:hist}
\end{figure}
\subsection{Full ReLaX Interpretability Results}
Next, we apply our retinal layer attribution method to produce quantitative explanations.  \Cref{tab:seg2} shows the $R_i$ (\Cref{eq:Fi}) values for correct classifications of each type. In all classifications, the ONL-ISM layer is the most focused followed by the NFL-IPL layer. This is likely due to these two regions occupying the most space in the scans. The rest of the layers have some variance between the diseases. CNV has the highest $R_i$ values for INL, OPL, and OS-RPE. DME has the highest $R_i$ for ONL-ISM. Drusen has the highest $R_i$ for ILM and NFL-IPL. Normal has the highest $R_i$ for ISE. In a sense, the interpretability of the model is similar to the reasoning of a human eye care professional. A human eye care professional looks at certain regions in the scan to make a decision, some regions more important than others. Our model, unlike previous ones, can give the same level of explanation, highlighting which regions are important for the classification. The added interpretability from ReLaX gives newfound quantitative model explanations as well as richer qualitative explanations. The standalone GradCAM heatmaps in \Cref{fig:seg1} only represent the rough regions where the model focused. ReLaX can name the specific retinal layers of focus and provide a quantitative measure. 

\begin{table}[t]
    \centering
    \caption{The model's focus during its two misclassifications. The first misclassification is the model predicting Normal on a DME scan. The columns labeled Difference represent the numeric difference between the retinal layer involvements during the misclassifications and the mean values for correct classifications. The first deviation column represents how many standard deviations the values in the first column are from the mean when correctly classifying Normal scans. The second misclassification is the model predicting CNV instead of Drusen. The second deviation column represents how many standard deviations the values in the third column are from the mean when correctly classifying CNV scans. (E.g. Misclassification 1 has 16.76\% focus on ILM, which is 8.80 greater than the mean for correct normal classification and 5.82 standard deviations from the mean)}
 
    \begin{tabular}{ccccccc}
        \toprule
        Layer   & Misclass. 1 & Difference & Deviation & Misclass. 2 & Difference & Deviation \\\midrule
        ILM     & 16.76\%     & 8.80\%      & 5.82      & 9.95\%    & 4.56\%         & 1.74      \\
        NFL-IPL & 18.67\%     & -1.81\%     & -0.60     & 21.43\%   & 7.40\%         & 1.66      \\
        INL     & 14.54\%     & 5.79\%      & 3.01      & 11.96\%   & -0.30\%         & -0.07     \\
        OPL     & 7.94\%      & -2.02\%     & -1.66     & 12.20\%   & -0.53\%         & -0.14     \\
        ONL-ISM & 33.52\%     & 2.61\%      & 0.85      & 28.00\%   & -7.91\%         & -1.14     \\
        ISE     & 3.76\%      & -9.42\%     & -5.20     & 9.05\%    & -0.86\%         & -0.31     \\
        OS-RPE  & 4.80\%      & -3.95\%     & -3.07     & 7.41\%    & -2.37\%         & -0.79     \\
        \bottomrule
    \end{tabular}

    \label{tab:seg5}
\end{table}

\subsection{Connection to Human Eyecare Professionals}
Now we demonstrate that our model is learning the correct areas of the scan to focus on. CNV involves the growth of new blood vessels from the choroid that result in breaking the Bruch's membrane and the subretinal space. This is consistent with our model which says that CNV has the highest focus of all diseases on the OS-RPE, the lowest region of the retina. DME occurs with excess fluid build up in the macula of the eye, located at the center of the retina. Again, this is consistent with our model which says that DME has the highest focus of all diseases on the ONL-ISM, the layer in the middle of the retina. Drusen is an accumulation of extracellular material which cause elevated RPE. While Drusen has low focus for the ISE and OS-RPE, it has the highest focus on the two regions at the top. This is likely because the model analyzes the peaks of the elevated RPE "humps" which occur closely to the top of the OCT scan. Finally, it makes sense that Normal does not have the most focus on many regions. This is because the model must first rule out the three diseases before making a Normal classification, similar to a human eye care professional. 

\subsection{Error Analysis}
When correctly classifying scans, \Cref{fig:hist} shows that retinal layer focus largely follow a Gaussian distribution. From a quick visual inspection, a significant majority of the 28 histograms follow a Gaussian distribution. The remaining histograms generally still have peaks close to the mean and thinner ends. This shows that obtaining retinal layer involvement percentages further from the mean (\Cref{tab:seg2}) implies a lower chance at a correct prediction. The chance of correct prediction is highest when the retinal layer involvements do not significantly deviate from the mean. In addition, \Cref{fig:hist} demonstrates that the segmentation errors (\Cref{fig:seg1}) do not significantly impede the quality of explanations. Even though some scans will have segmentation errors, their retinal layer involvements still fall closely to the mean. Using this insight on the distribution of $R_i$ values, we perform error analyses on the model's two misclassifications.   

\begin{figure}[t]
    \centering
    \includegraphics[width=\linewidth]{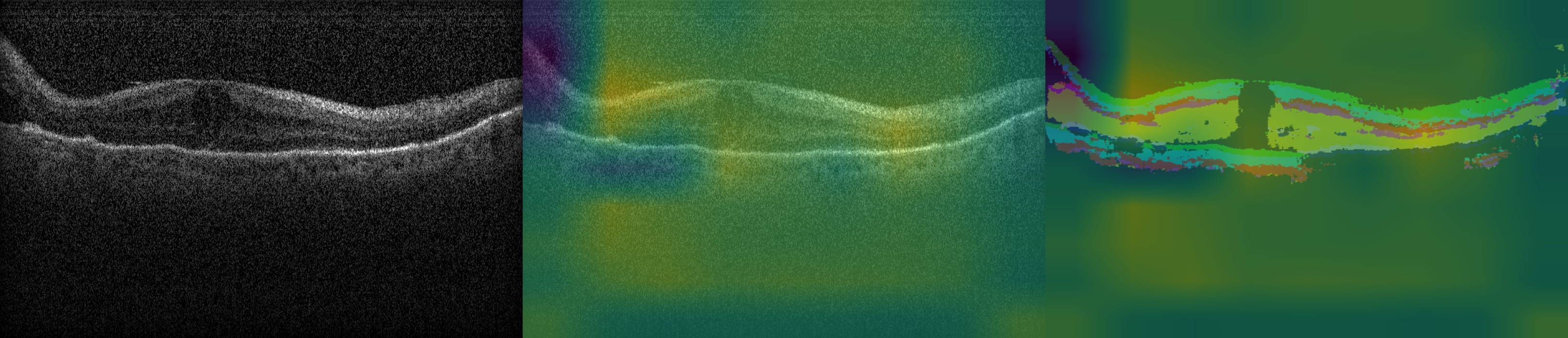}
    \includegraphics[width=\linewidth]{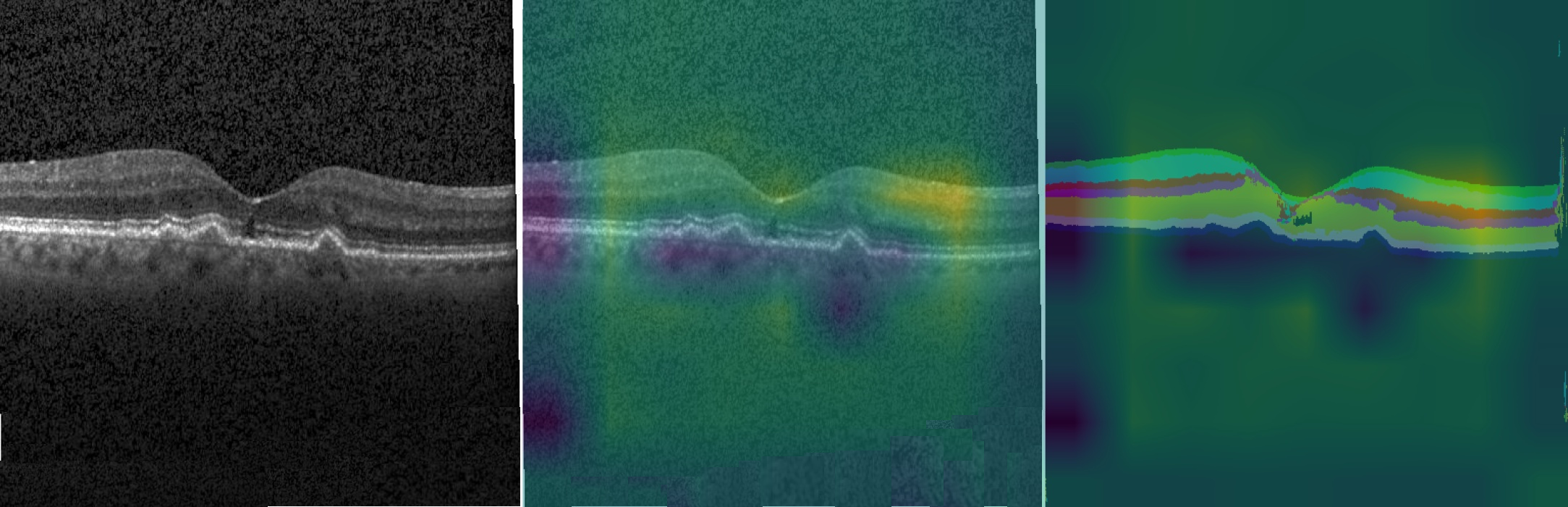}
    \caption{The two misclassifications from our model. Left: OCT scan. Middle: OCT scan with heatmap overlayed, Right: Segmentation map with heatmap overlayed. The first row is a DME scan which the model predicted Normal. The second row is a Drusen scan which the model predicted CNV.}
    \label{fig:seg4}
\end{figure}

\Cref{fig:seg4} and \Cref{tab:seg5} explain the model's incorrect classifications. The GradCAM overlayed heatmaps show the model has no specific area of focus in both classifications. The model's focus is nearly evenly distributed across the entire OCT scan instead of a centralized area commonly seen during correct classifications. The deviations from the second column of \Cref{tab:seg5} show the ILM, INL, ISE, and OS-RPE are significantly off from the mean $R_i$ for correct Normal classifications. The large deviations show that the model's explanation is too different from the explanation of a correct classification. As shown in \Cref{fig:hist}, larger deviations directly correlate to lower chances at correct predictions. When a prediction has a statistically suspicious explanation, the scan should be flagged for further review; the model's prediction should not be trusted. 

We analyze the probability of correct classification by finding the difference values in the histograms of \Cref{fig:hist}. An 8.80 difference in ILM involvement for correct normal classification is way off the right end of the histogram. Similarly differences of 5.79, -9.42, and -3.95 for INL, ISE, and OS-RPE fall very far on the ends of the histogram. This shows the low probability of these classifications being correct. 

On the other hand, the other set of deviations from the DME scan mistaken CNV appear to be much closer. This is due to the correct CNV classifications generally having higher variance in their $R_i$. While this example is not as drastic as the other misclassification, there are still some areas of uncertainty. The ILM and NFL-IPL involvements are quite higher than the mean. These difference values correspond to areas of the histogram (\Cref{fig:hist} Rows 1 and 2, Column 1) which are much less populated. In a sample size of 242 correct CNV classifications, less than twenty scans have higher ILM deviations. There is a similar story for the NFL-IPL layer involvement. The rest of the layer involvements are relatively close to the mean signaling a somewhat valid interpretation. While the model appears to have a statistically sound quantitative explanation for this classification, the qualitative explanation in the heatmap shows the model is not confident when making this prediction. 

\section{Discussion} 

Making deep learning suitable for clinical application requires significant progress in explainability. Accurate classification of OCT is a key task for ophthalmic clinics; automation could significantly alleviate the current workload. In previous works for OCT classification, attribution based methods were the best form of model interpretations. However, attribution methods such as occlusion mapping or CAM, fall short of providing the same level of explanation as a human doctor. While deep learning has already shown its potential in classifying OCT, a more in depth form of model explanation could allow full integration of neural networks into the clinical workflow. 

Our primary approach to generating greater insight into the classification model is the concept of \textbf{retinal layer attribution}. We rethink attribution from a layer-by-layer perspective rather than a pixel-by-pixel perspective. Through our analysis of ReLaX, we find that our classification models have a strong correlation to human judgement. For each disease, the model knows which retinal layer is important for making the correct classification. We verify that human eye care professionals also use the same retinal layers to make diagnoses. This level of interpretability is much stronger than the previous works who simply have human experts verify heatmaps. Having humans verify a heatmap does not fully ensure that the model is behaving correctly. Heatmaps can be very general and humans can be subjective. Our work specifically pinpoints the retinal layers in question and verifies the model's behaviours. Furthermore, ReLaX can also be used for insightful error analysis of the model. \Cref{tab:seg5} and \Cref{fig:seg4} show noticeable indicators that the predictions generated are incorrect. To our knowledge, our work is the first to perform significant error analysis. In a clinical setting, error analysis is very important for maximizing confidence in diagnoses.  

We also contribute a state-of-the-art OCT classification model. As shown in \Cref{Tab:metrics}, our model performs more accurately than the previously developed OCT classification models on the Kermany OCT data set. The model we contribute only makes two misclassifications on the entire data set. When paired with a thorough system for error analysis, this model can be clinically implemented with a high degree of confidence. In the cases where the model makes a mistake, it does not go undetected due to the error analysis capabilities of ReLaX. 

The main limitations in our work are the lack of publicly-available data for OCT related tasks. We utilize the Kermany OCT data set, but the data set is not sufficiently competitive. While we do contribute a higher accuracy than previous works, several others have developed highly accurate OCT classification models on the data set. The consequence of the noncompetitive data set is that we do not have many incorrect samples to evaluate our proposed ReLaX method. In addition to the lack of available OCT classification data, there is also a lack of ground-truth segmentations for OCT scans. The Kermany data set we utilize does not provide ground truths, thus it is difficult for us to evaluate and improve our segmentation model. While we show that this does not largely impact the efficacy of our algorithm (\Cref{fig:hist}), accurate segmentation would provide more certainty. 

In the future, more publicly-available OCT data sets must be released for the further development of deep learning in ophthalmology. We hope that we can evaluate our approach on a more competitive OCT data set. The Kermany OCT data set has become insufficient for future work as it is possible to achieve nearly perfect accuracy. Furthermore, we wish to improve the accuracy of our segmentation models through employing newer techniques such as the one provided by He et al.~\cite{he_fully_2019}. However, our method will require a competitive data set with both ground-truths for OCT classification and OCT segmentation. Another possible direction could be to further strengthen the claims that humans and deep learning models are alike through an eye-tracker study of human professionals analyzing OCT scans.  
\section{Conclusion}
In this work, we present ReLaX, a novel retinal layer attribution algorithm for producing rich explanations of OCT classification. We also develop a highly accurate, state-of-the-art model for classifying OCT scans of CNV, DME, Drusen, and Normal. Our classification model performs at an accuracy of 99.8\%, higher than all previous works. ReLaX utilizes a novel combination of GradCAM heatmaps and segmented OCT scans to accurately pinpoint the retinal layers important for classification. Our work rethinks attribution-based explanations by focusing on the importance of certain regions in a classification rather than of each pixel. Pixel-based attribution does not accurately demonstrate the important ideas behind a model's decision. Retinal layer attribution can give more-detailed qualitative explanations as well as novel quantitative explanations. ReLaX confirms the similarities between human and computer-aided analysis of retinal imaging. When patients believe that they can get the same level of care from a human and a computer, they are more likely to trust a computer-aided diagnosis.Furthermore, ReLaX gives insightful analytics on the model's likelihood of being correct for a given prediction. Through our investigation, we find that incorrect predictions by the model can often be detected early. In a clinical setting, this prevents false diagnoses which can lead to improper plans of treatment. This new level of interpretability in our model brings deep learning one step closer to real clinical application; patients and doctors can be more confident that they are getting a correct diagnosis. In addition to an accurate diagnosis, patients deserve qualitative and quantitative information as to why the model made such decision rather than a standalone diagnosis. Only then will patients be able to relax.

%\clearpage\mbox{}Page \thepage\ of the manuscript.
%\clearpage\mbox{}Page \thepage\ of the manuscript.

%This is the last page of the manuscript.
%\par\vfill\par
%Now we have reached the maximum size of the ECCV 2022 submission (excluding references).
%References should start immediately after the main text, but can continue on p.15 if needed.

\clearpage
\bibliographystyle{splncs04}
\bibliography{egbib}

\begin{thebibliography}{10}
\providecommand{\url}[1]{\texttt{#1}}
\providecommand{\urlprefix}{URL }
\providecommand{\doi}[1]{https://doi.org/#1}

\bibitem{brightfocusfoundation_2020}
Age-related macular degeneration: Facts \& figures (Jan 2020),
  \url{https://www.brightfocus.org/macular/article/age-related-macular-facts-figures}

\bibitem{asif_amjad_qurrat-ul-ain_2022}
Asif, S., Amjad, K., ul~Ain, Q.: Deep residual network for diagnosis of retinal
  diseases using optical coherence tomography images - interdisciplinary
  sciences: Computational life sciences. SpringerLink  (Jun 2022),
  \url{https://link.springer.com/article/10.1007/s12539-022-00533-z\#citeas}

\bibitem{de_fauw}
De~Fauw, J., et~al.: Clinically applicable deep learning for diagnosis and
  referral in retinal disease (Aug 2018),
  \url{https://www.nature.com/articles/s41591-018-0107-6/}

\bibitem{deng2009imagenet}
Deng, J., Dong, W., Socher, R., Li, L.J., Li, K., Fei-Fei, L.: Imagenet: A
  large-scale hierarchical image database. In: 2009 IEEE conference on computer
  vision and pattern recognition. pp. 248--255. Ieee (2009)

\bibitem{DBLP:journals/corr/abs-2010-11929}
Dosovitskiy, A., Beyer, L., Kolesnikov, A., Weissenborn, D., Zhai, X.,
  Unterthiner, T., Dehghani, M., Minderer, M., Heigold, G., Gelly, S.,
  Uszkoreit, J., Houlsby, N.: An image is worth 16x16 words: Transformers for
  image recognition at scale. CoRR  \textbf{abs/2010.11929} (2020),
  \url{https://arxiv.org/abs/2010.11929}

\bibitem{pmid28128796}
Foot, B., MacEwen, C.: {{S}urveillance of sight loss due to delay in ophthalmic
  treatment or review: frequency, cause and outcome}. Eye (Lond)
  \textbf{31}(5),  771--775 (May 2017)

\bibitem{pmid27409459}
Fujimoto, J., Swanson, E.: {{T}he {D}evelopment, {C}ommercialization, and
  {I}mpact of {O}ptical {C}oherence {T}omography}. Invest Ophthalmol Vis Sci
  \textbf{57}(9),  OCT1--OCT13 (07 2016)

\bibitem{pmid10933065}
Fujimoto, J.G., Pitris, C., Boppart, S.A., Brezinski, M.E.: {{O}ptical
  coherence tomography: an emerging technology for biomedical imaging and
  optical biopsy}. Neoplasia  \textbf{2}(1-2),  9--25 (2000)

\bibitem{he_fully_2019}
He, Y., Carass, A., Liu, Y., Jedynak, B.M., Solomon, S.D., Saidha, S.,
  Calabresi, P.A., Prince, J.L.: Fully {Convolutional} {Boundary} {Regression}
  for {Retina} {OCT} {Segmentation}. In: Shen, D., Liu, T., Peters, T.M.,
  Staib, L.H., Essert, C., Zhou, S., Yap, P.T., Khan, A. (eds.) Medical {Image}
  {Computing} and {Computer} {Assisted} {Intervention} – {MICCAI} 2019. pp.
  120--128. Lecture {Notes} in {Computer} {Science}, Springer International
  Publishing, Cham (2019). \doi{10.1007/978-3-030-32239-7\_14}

\bibitem{kermany2018identifying}
Kermany, D.S., et~al.: Identifying medical diagnoses and treatable diseases by
  image-based deep learning. Cell  \textbf{172}(5),  1122--1131 (2018)

\bibitem{Kingma2015AdamAM}
Kingma, D.P., Ba, J.: Adam: A method for stochastic optimization. CoRR
  \textbf{abs/1412.6980} (2015)

\bibitem{NIPS2012_4824}
Krizhevsky, A., Sutskever, I., Hinton, G.E.: Imagenet classification with deep
  convolutional neural networks. In: Pereira, F., Burges, C.J.C., Bottou, L.,
  Weinberger, K.Q. (eds.) Advances in Neural Information Processing Systems 25,
  pp. 1097--1105. Curran Associates, Inc. (2012)

\bibitem{pmid31853395}
Li, F., et~al.: {{D}eep learning-based automated detection of retinal diseases
  using optical coherence tomography images}. Biomed Opt Express
  \textbf{10}(12),  6204--6226 (Dec 2019)

\bibitem{10.1167/tvst.7.6.41}
Lu, W., Tong, Y., Yu, Y., Xing, Y., Chen, C., Shen, Y.: {Deep Learning-Based
  Automated Classification of Multi-Categorical Abnormalities From Optical
  Coherence Tomography Images}. Translational Vision Science \& Technology
  \textbf{7}(6),  41--41 (12 2018). \doi{10.1167/tvst.7.6.41},
  \url{https://doi.org/10.1167/tvst.7.6.41}

\bibitem{pmid31697697}
Nagasato, D., et~al.: {{A}utomated detection of a nonperfusion area caused by
  retinal vein occlusion in optical coherence tomography angiography images
  using deep learning}. PLoS One  \textbf{14}(11),  e0223965 (2019)

\bibitem{puneet_optical_2022}
{Puneet}, Kumar, R., Gupta, M.: Optical coherence tomography image based eye
  disease detection using deep convolutional neural network. Health Information
  Science and Systems  \textbf{10}(1), ~13 (Jun 2022).
  \doi{10.1007/s13755-022-00182-y},
  \url{https://doi.org/10.1007/s13755-022-00182-y}

\bibitem{DBLP:journals/corr/RonnebergerFB15}
Ronneberger, O., Fischer, P., Brox, T.: U-net: Convolutional networks for
  biomedical image segmentation. CoRR  \textbf{abs/1505.04597} (2015),
  \url{http://arxiv.org/abs/1505.04597}

\bibitem{DBLP:journals/corr/RoyCKSKWN17}
Roy, A.G., et~al.: Relaynet: Retinal layer and fluid segmentation of macular
  optical coherence tomography using fully convolutional network. CoRR
  \textbf{abs/1704.02161} (2017), \url{http://arxiv.org/abs/1704.02161}

\bibitem{saleh_transfer_2022}
Saleh, N., Abdel~Wahed, M., Salaheldin, A.M.: Transfer learning-based platform
  for detecting multi-classification retinal disorders using optical coherence
  tomography images. International Journal of Imaging Systems and Technology
  \textbf{32}(3),  740--752 (2022). \doi{10.1002/ima.22673},
  \url{https://onlinelibrary.wiley.com/doi/abs/10.1002/ima.22673}, \_eprint:
  https://onlinelibrary.wiley.com/doi/pdf/10.1002/ima.22673

\bibitem{DBLP:journals/corr/SelvarajuDVCPB16}
Selvaraju, R.R., Das, A., Vedantam, R., Cogswell, M., Parikh, D., Batra, D.:
  Grad-cam: Why did you say that? visual explanations from deep networks via
  gradient-based localization. CoRR  \textbf{abs/1610.02391} (2016),
  \url{http://arxiv.org/abs/1610.02391}

\bibitem{simonyan_vedaldi_zisserman_2014}
Simonyan, K., Vedaldi, A., Zisserman, A.: Deep inside convolutional networks:
  Visualising image classification models and saliency maps (Apr 2014),
  \url{https://arxiv.org/abs/1312.6034}

\bibitem{1412.6806}
Springenberg, J.T., Dosovitskiy, A., Brox, T., Riedmiller, M.: Striving for
  simplicity: The all convolutional net (2014)

\bibitem{Srinivasan:14}
Srinivasan, P.P., et~al.: Fully automated detection of diabetic macular edema
  and dry age-related macular degeneration from optical coherence tomography
  images. Biomed. Opt. Express  \textbf{5}(10),  3568--3577 (Oct 2014).
  \doi{10.1364/BOE.5.003568},
  \url{http://www.osapublishing.org/boe/abstract.cfm?URI=boe-5-10-3568}

\bibitem{DBLP:journals/corr/abs-1905-11946}
Tan, M., Le, Q.V.: Efficientnet: Rethinking model scaling for convolutional
  neural networks. CoRR  \textbf{abs/1905.11946} (2019),
  \url{http://arxiv.org/abs/1905.11946}

\bibitem{pmid32192460}
Tsuji, T., et~al.: {{C}lassification of optical coherence tomography images
  using a capsule network}. BMC Ophthalmol  \textbf{20}(1), ~114 (Mar 2020)

\bibitem{8794616}
Wang, D., Wang, L.: On oct image classification via deep learning. IEEE
  Photonics Journal  \textbf{11}(5),  1--14 (2019).
  \doi{10.1109/JPHOT.2019.2934484}

\bibitem{pmid33492598}
Yoo, T.K., Choi, J.Y., Kim, H.K.: {{F}easibility study to improve deep learning
  in {O}{C}{T} diagnosis of rare retinal diseases with few-shot
  classification}. Med Biol Eng Comput  \textbf{59}(2),  401--415 (Feb 2021)

\bibitem{DBLP:journals/corr/ZeilerF13}
Zeiler, M.D., Fergus, R.: Visualizing and understanding convolutional networks.
  CoRR  \textbf{abs/1311.2901} (2013), \url{http://arxiv.org/abs/1311.2901}

\bibitem{zhou2015cnnlocalization}
Zhou, B., Khosla, A., A., L., Oliva, A., Torralba, A.: {Learning Deep Features
  for Discriminative Localization.} CVPR  (2016)

\end{thebibliography}
\end{document}